\documentclass[12pt,preprint]{aastex}
\usepackage{graphicx}
\usepackage{amsmath}

\begin{document}
\title{Polarization of Quasars: Electronic Scattering in the Broad Absorption Line Region}
\author{ Hui-Yuan Wang, Ting-Gui Wang and Jun-Xian Wang}
\date{}
\email{whywang@mail.ustc.edu.cn} \affil{Center for
Astrophysics,University of Science and Technology of China, Hefei,
230026, China}

\begin{abstract}
It is widely accepted that the broad absoption line region (BALR)
exists in most (if not all) quasars with a small covering factor.
Recent works showed that the BALR is optically thick to soft and
even medium energy X-rays, with a typical hydrogen column density
of a few 10$^{23}$ to $>$ 10$^{24}$ cm$^{-2}$. The electronic
scattering in the thick absorber might contribute significantly to
the observed continuum polarization for both BAL QSOs and non-BAL
QSOs. In this paper, we present a detailed study of the electronic
scattering in the BALR by assuming an equatorial and axisymmetric
outflow model. Monte-Carlo simulations are performed to correct
the effect of and radiation transfer and attenuation. Assuming an
average covering factor of 0.2 of the BALR, which is consistent
with observations, we find the electronic scattering in the BALR
with a column density of $\sim$ 4 $\times$ 10$^{23}$ cm$^{-2}$ can
successfully produce the observed average cotinuum polarization
for both BAL QSOs and non-BAL QSOs. The observed distribution of
the continuum polarization of radio quiet quasars (for both BAL
QSOs and non-BAL QSOs) is helpful to study the dispersal
distribution of the BALR. We find that, to match the observations,
the maximum continuum polarization produced by the BALR (while
viewed edge-on) peaks at $P$ = 0.34\%, which is much smaller than
the average continuum polarization of BAL QSOs ($P$ = 0.93\%). The
discrepancy can be explained by a selection bias, that the BAL
with larger covering factor, and thus producing larger continuum
polarization, is more likely to be detected. A larger sample of
radio quiet quasars with accurate measurement of the continuum
polarization will help give better constraints to the distribution
of the BALR properties.
\end{abstract}

\keywords {polarization-scattering-quasars: absorption lines}

\section{Introduction}

About 10-20 \%  optically selected QSOs exhibit broad absorption
troughs in resonant lines up to 0.1c blueward of the corresponding
emission lines (Hewett \& Foltz 2003, Reichard et al. 2003).
Usually the Broad Absorption Lines (BAL) are detected only in high
ionization ones, such as CIV, NV, SiIV and OVI, named high ionized
BAL (HiBAL), but 10\% of BAL QSOs show also low ionization lines
(LiBAL), such as MgII, AlIII and even FeII. The blue shift of the
absorption lines suggests that they are formed in a partially
ionized wind, outflowing from the quasar. The observed flux ratios
of the emission to absorption line imply that the covering factor
of the BAL Region (BALR)  be $<$ 20\% (Hamann, Korista \& Morris
1993). Futhermore, the properties of UV/X-ray continuum and
emission lines of BAL QSOs and non-BAL QSOs are also found to be
similar (Weymann et al. 1991; Reichard et al. 2003; Green et al.
2001). These facts lead to a general picture that BALR covers only
a small fraction of sky and may be present in every quasar
(Weymann et al. 1991). We note there are evidences suggesting that
LiBAL QSOs are in a special evolution phase of quasars rather than
merely viewed on a special inclination (Voit, Weymann \& Korista
1993; Canalizo \& Stockton 2001; c.f., Willott, Rawlings,\& Grimes
2003; Lewis, Chapman, \& Kuncic 2003).

The evidence for non-spherical BALR is also supported by the
discovery that BAL QSOs are usually more polarized than non-BAL
QSOs (Ogle et al. 1999). BAL QSO is the only high polarization
population among the radio quiet QSOs (e.g., Stockman, Moore \&
Angel 1984). Stockman et al. found 9 of 30 BAL QSOs show high
polarization ($P>1.5\%$) in the optical continuum. The result was
confirmed later by Schmidt \& Hines (1999), who found the average
polarization degree of BAL QSOs is 2.4 times that of the optically
selected quasars. Consistently, Hutsem$\grave{e}$kers \& Lamy
(2001) obtained average polarization degrees of $\sim$ 0.43$\%$,
0.93$\%$ and 1.46$\%$ for non-BAL QSOs, HiBAL QSOs and LiBAL QSOs
respectively. This result is similar to that of Schmidt \& Hines
(1999): 0.4$\%$ and 1.0$\%$ for non-BAL and BAL QSOs. It is also
evident that the polarization distribution of non-BAL QSOs drops
sharply toward high polarization (see also Berriman et al. 1990).

BAL QSOs are notorious X-ray weak following the work of Green \&
Mathur (1996; see also Brinkmann et al. 1999). Now we have strong
evidences that the weakness in X-rays is not intrinsic but due to
strong X-ray absorption, with the hydrogen column density of a few
10$^{23}$ to $>$ 10$^{24}$~cm$^{-2}$ (Wang et al. 1999; Gallagher
et al. 2000; Mathur et al. 2000).  The X-ray absorber might be
responsible for the recently detected blueshifted X-ray BALs,
suggesting they are outflowing at even higher velocities and
highly ionized (Chartas et al. 2002; 2003). The electronic
scattering in the thick X-ray absorber may contribute
significantly to the observed continuum polarization for both BAL
QSOs and non-BAL QOSs, if BALR exists in every quasar. In this
paper, we perform detailed calculations and Monte Carlo
simulations to study the polarization produced by electronic
scattering in the BALR. By comparing the expected polarizations
with the observed ones, we give strong contraints to the BALR
model.

\section{Models and the Monte-Carlo method}

There are many dynamic models for the BAL outflow, depending on
the flow type (hydrodynamic or hydromagnetic flow) and the
accelerating mechanisms (radiation, gas pressure or magnetic
field). In most models, the flow is accelerated through the
resonant line scattering, which is supported by the line-locking
phenomena. A general difficulty of such models, however, is to
prevent the flow from being fully ionized while its average
density drops rapidly with increasing radius. One solution
proposed by Murry \& Chiang (1995, hereafter MC95) is that the
flow is shielded from the intense soft X-ray radiation by highly
ionized medium in the inner region with a typical column density
of a few $10^{23}$ cm$^{-2}$ to 10$^{24}$~cm$^{-2}$. This
shielding gas can also account for the observed heavy X-ray
absorption. This model is in qualitative agreement with more recent
hydrodynamic calculation of the radiative accelerated wind from an
accretion disk (Proga, Stone \& Kallman. 2001). The second
solution is the two-phase flow, in which a dense, low ionization,
cold clouds are embeded in a highly ionized, hot and tenuous medium.
The cold clouds with a small filling factor, accelerated by the
line and continuum radiation pressure to high velocities, is
responsible for the BAL features. An implement to the second
scheme is the massive hydromagnetic and radiative driven wind
model. In the model, the massive high-ionized continuous outflow
is driven centrifugally, and accelerated radiatively by the
central continuum source (Everett 2002; Konigl \& Kartje 1994; de
Kool \& Begelman 1995). The total column density of the hot medium
is also very large (with N$_H=10^{22}$ cm$^{-2}\sim 10^{26}$
cm$^{-2}$ ). In fact such a medium itself can also be considered
as the shielding gas. A third scheme is the dusty wind due to the
mass loss of stars in the nucleus; both the dust absorption and
the line scattering contribute to the accelerating force of the
gas (Voit et al. 1993; Scoville \& Norman 1995).

\begin{figure}
  \plotone{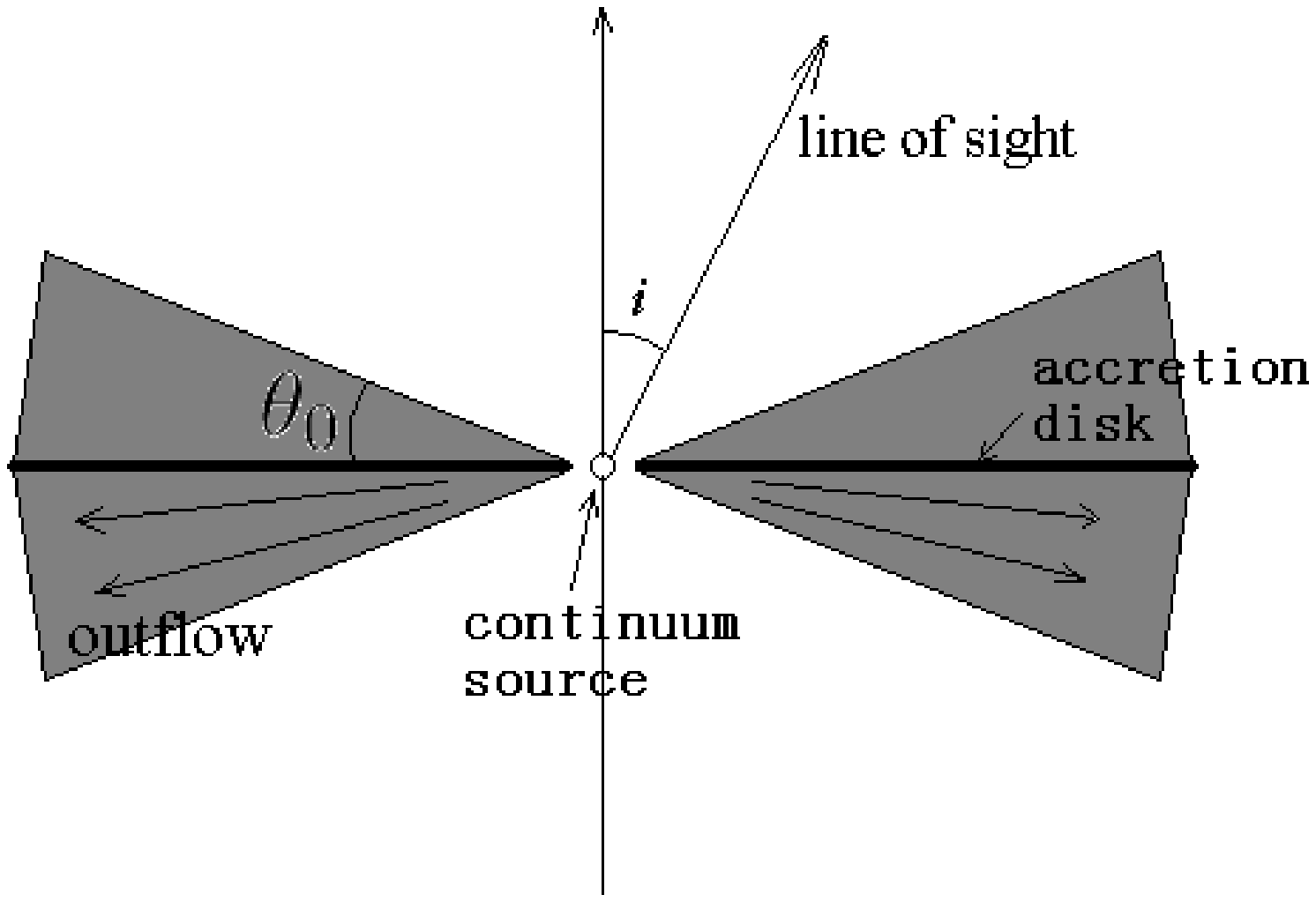}
  \caption{The geometry of the outflow model used in the paper.
   } \label{model}
\end{figure}

Following MC95, we assume an equatorial and axisymmetric outflow,
with a half open angle $\theta_0$ (see Fig. \ref{model}). In this
paper, we focus on the continuum polarization produced by the
electronic scattering in the outflow, and leave the study of the
resonant scattering, which can produce obvious polarization around
the broad absorption trough, in a future paper (Wang et al. in
prep). The shielding gas is the major source of the electronic
scattering and X-ray photo-electronic absorption. Since both the
scattering and X-ray absorption are insensitive to the density
profile of the shielding gas, we assume a constant electron
density in this region. We consider a range of column densities
($10^{23}$ to $>10^{24}$~cm$^{-2}$) for this region, consistent
with that obtained from the X-ray observations.

Following Lee, Blandford \& Western (1994), we denote the density
of the polarized incident photons in direction
$(\theta_i,\varphi_i)$ as

\begin{equation}
\left( \begin{array}{cc}
    \rho_{11}^i &\rho_{12}^i \nonumber \\
    \rho_{21}^i&\rho_{22}^i
    \end{array} \right)
\end{equation}

where $\rho_{ij}$ is the i,j component of the photon-density
matrix. The outward density in the direction
$(\theta_o,\varphi_o)$ is

\begin{equation}
\left( \begin{array}{cc}
  \rho_{11}^o & \rho_{12}^o \nonumber \\
  \rho_{21}^o &\rho_{22}^o
\end{array} \right).
\end{equation}

For the electronic scattering we may write the outward photon
density after one scattering as (see Chandrasekar 1950):

\begin{eqnarray}
\rho_{11}^o & \propto &
\rho_{11}^i[\cos^2\theta_o\cos^2\theta_i\cos^2(\varphi_i-
     \varphi_o)+2\cos\theta_o\cos\theta_i\sin\theta_o\sin\theta_i\cos
     (\varphi_i-\varphi_o)+\sin^2\theta_o\sin^2\theta_i] \nonumber\\
 & &   +\rho_{22}^i\cos^2\theta_o\sin^2(\varphi_i-\varphi_o)+\frac{1}{2}(\rho_{12}^i+\rho_{21}^i)
       [\cos^2\theta_o\cos\theta_i\sin(2\varphi_i-2\varphi_o) \nonumber\\
&
&+2\cos\theta_o\sin\theta_o\sin\theta_i\sin(\varphi_i-\varphi_o)]\label{eq:rho11}
\end{eqnarray}
\begin{equation}\label{eq:rho22}
\rho_{22}^o\propto\rho_{11}^i\cos^2\theta_i\sin^2(\varphi_i-\varphi_o)+
\rho_{22}^i\cos^2(\varphi_i-\varphi_o)-\frac{1}{2}(\rho_{12}^i+\rho_{21}^i)\cos\theta_i\sin(2\varphi_i-2\varphi_o)
\end{equation}
\begin{eqnarray}\label{eq:rho12}
\rho_{12}^o & \propto &
\rho_{11}^i[-\frac{1}{2}\cos\theta_o\cos^2\theta_i
\sin(2\varphi_i-2\varphi_o)-\sin\theta_o\sin\theta_i\cos\theta_i\sin(\varphi_i-
\varphi_o)] \nonumber\\
&
&+\frac{1}{2}\rho_{22}^i\cos\theta_o\sin(2\varphi_i-2\varphi_o)+\rho_{21}^i[-\frac{1}{2}\cos\theta_o\cos\theta_i+\frac{1}{2}\cos\theta_o
\cos\theta_i\cos(2\varphi_i-2\varphi_o)]\nonumber\\
&
&+\rho_{12}^i[\frac{1}{2}\cos\theta_o\cos\theta_i+\frac{1}{2}\cos\theta_o
\cos\theta_i\cos(2\varphi_i-2\varphi_o)+\sin\theta_o\sin\theta_i\cos(\varphi_i-\varphi_o)]
\end{eqnarray}
\begin{eqnarray}\label{eq:rho21}
\rho_{21}^o& \propto
&\rho_{11}^i[-\frac{1}{2}\cos\theta_o\cos^2\theta_i\sin
(2\varphi_i-2\varphi_o)-\sin\theta_o\sin\theta_i\cos\theta_i\sin(\varphi_i-
\varphi_o)]\nonumber\\ &
&+\frac{1}{2}\rho_{22}^i\cos\theta_o\sin(2\varphi_i-2\varphi_o)
 +\rho_{12}^i[-\frac{1}{2}\cos\theta_o\cos\theta_i+\frac{1}{2}\cos\theta_o
\cos\theta_i\cos(2\varphi_i-2\varphi_o)] \nonumber\\
&
&+\rho_{21}^i[\frac{1}{2}\cos\theta_o\cos\theta_i+\frac{1}{2}\cos\theta_o
\cos\theta_i\cos(2\varphi_i-2\varphi_o)+\sin\theta_o\sin\theta_i\cos(\varphi_i
-\varphi_o)]
\end{eqnarray}

The four STOKES parameters read,

\begin{equation}
I=\rho_{11}+\rho_{22},\;\;
Q=\rho_{11}-\rho_{22},\;\;U=\rho_{12}+\rho_{21},\;\;V=\rho_{12}-\rho_{21}
\end{equation}

Other interesting quantities, such as the total flux, the PA
rotation, the polarization degree and the polarized flux, can be
calculated from the Stokes parameters. The polarization degree
follows

\begin{equation}
  p=\frac{\sqrt{Q^2+U^2}}{I}
\end{equation}

Following the steps below, we run Monte-Carlo simulations to
calculate the output Stokes parameters for given incident
radiation and spatial distribution of the scatterer.

\begin{itemize}
\item For each incident photon, we give a random propagation
direction. Then, the unit density matrix of the photon with a
given polarization degree is calculated (Lee 1994). \item The
probability of the photon passing through an absorber with an
optical depth $\tau$ is  $p = \exp(-\tau)$. By assigning a random
number in [0,1] to $p$, we calculate the corresponding $\tau$. If
$\tau$ is larger than the real optical depth along the photon
propagation direction ($\tau'$), the photon will escape from the
medium and be collected in the output basket. \item If
$\tau'>\tau$, the photon will be scattered at $\vec{r}$, where the
real optical depth along the path of the photon is $\tau$. The
emergent direction and the density matrix of the scattered photon
is calculated (equations
\ref{eq:rho11},\ref{eq:rho22},\ref{eq:rho12} $\&$\ref{eq:rho21}).
\item The frequency of the scattered photon is calculated by
taking into consideration of Doppler shift, which is a function of
the incident direction, emergent direction and the velocity vector
of the scattering particle. The velocity vector of the scattering
particle is determined by the bulk velocity and the thermal
velocities at $\vec{r}$ (Lee $\&$ Blandford 1997).
\end{itemize}

We repeat step 2, 3, 4 until the photon either escapes or is
absorbed by the accretion disk, which is assumed to be optically
thick with no reflection.

\section{Results and Discussion}

In the case of optical thin limit ($\tau <<$ 1), the polarization
degree viewed at an inclination angle $i$ can be written
analytically as (Brown \& McLean 1978, hereafter BM78)

\begin{equation}
 P_e=2\bar{\tau}(1-3\gamma)\sin^2i\label{eq:p_e}
\end{equation}

where

\begin{equation}
\gamma =
\frac{\int_{-1}^{+1}\mu^2\tau(\mu)d\mu}{\int_{-1}^{+1}\tau(\mu)d\mu}
\end{equation}

 and

\begin{equation}
\bar{\tau}=\frac{3}{32}\int_{-1}^{+1}\tau(\mu)d\mu
\end{equation}

where $\mu = \cos\theta$ and $\tau(\mu)$ is the Thomson depth at
different polar angle $\theta$.

If BALR covers the inclination angle from 90$^{\rm o}$ -
$\theta_0$ to 90$^{\rm o}$ + $\theta_0$ (see Fig. \ref{model}) and
the accretion disk is optically thick with no reflection, the
average polarization due to the electron scattering for BAL and
non-BAL QSO can be written as

\begin{equation}
R\equiv
\frac{<P_e^{BAL}>}{<P_e^{non-BAL}>}=\frac{(1-\mu_0)(1-\mu_0^2/3)}{2/3-\mu_0+\mu_0^3/3}
\end{equation}

where $\mu_0=\sin\theta_0$.

Assuming a constant Thomson depth ($\tau_0$) of the the shielding
gas at different directions, we obtain the average polarization
for BAL QSOs

\begin{equation}
\bar{P}_{B}=(1-\frac{1}{3}\mu_0^2)\frac{3}{16}\tau_0\mu_0(1-\mu_0^2)
\end{equation}

The constant column density of the shielding gas at different
direction is over-simplified. However, a distribution of $\tau$
will not significantly change the estimated mean optical depth
($\int_{-\mu_0}^{\mu_0} \tau(\mu) d\mu$) if $\sin\theta_0$ is
small.

Using the observed values of R = 2.2 and $\bar{P}_{B}$ = 0.93\%
(see \S1), we obtain $\mu_0$=0.5 and $\tau_0$=0.144. $\mu_0=0.5$
is much larger than 0.2 derived from the observed fraction of BAL
QSOs. We point out the discrepancy might be due to selection bias,
if $\mu_0$ has a dispersal distribution instead of a $\delta$
function for quasars. We can see that we have higher chance to
detect BAL QSOs with higher $\mu_0$, and reversely, higher chance
for non BAL QSOs with lower $\mu_0$. Thus the average $\mu_0$ of
the observed BAL QSOs tends to be higher than that derived from
the fraction of BAL QSOs (also see Morris 1988), and that of the
observed non-BAL QSOs tends to be lower. For example, an average
$\mu_0$ = 0.25 for the observed BAL QSOs and $\mu_0$ = 0.188 for
the observed non-BAL QSOs can easily macth the observed value R =
2.2. Using $\mu_0 = 0.2$, to produce the observed average polarization
of BAL QSOs, $\tau_0=0.26$ or $N_e=3.9\times10^{23}$ cm$^{-2}$ is required.
If assuming an average $\mu_0 = 0.25$ for the observed BAL QSOs,
we obtained a slightly lower $N_e=3.25\times10^{23}$ cm$^{-2}$.
The predicted average column density is consistent with those
derived from the X-ray observations (Wang et al. 1999; Gallagher
et al. 2000; Mathur et al. 2000).

\begin{figure}
  \epsscale{1.0}\plotone{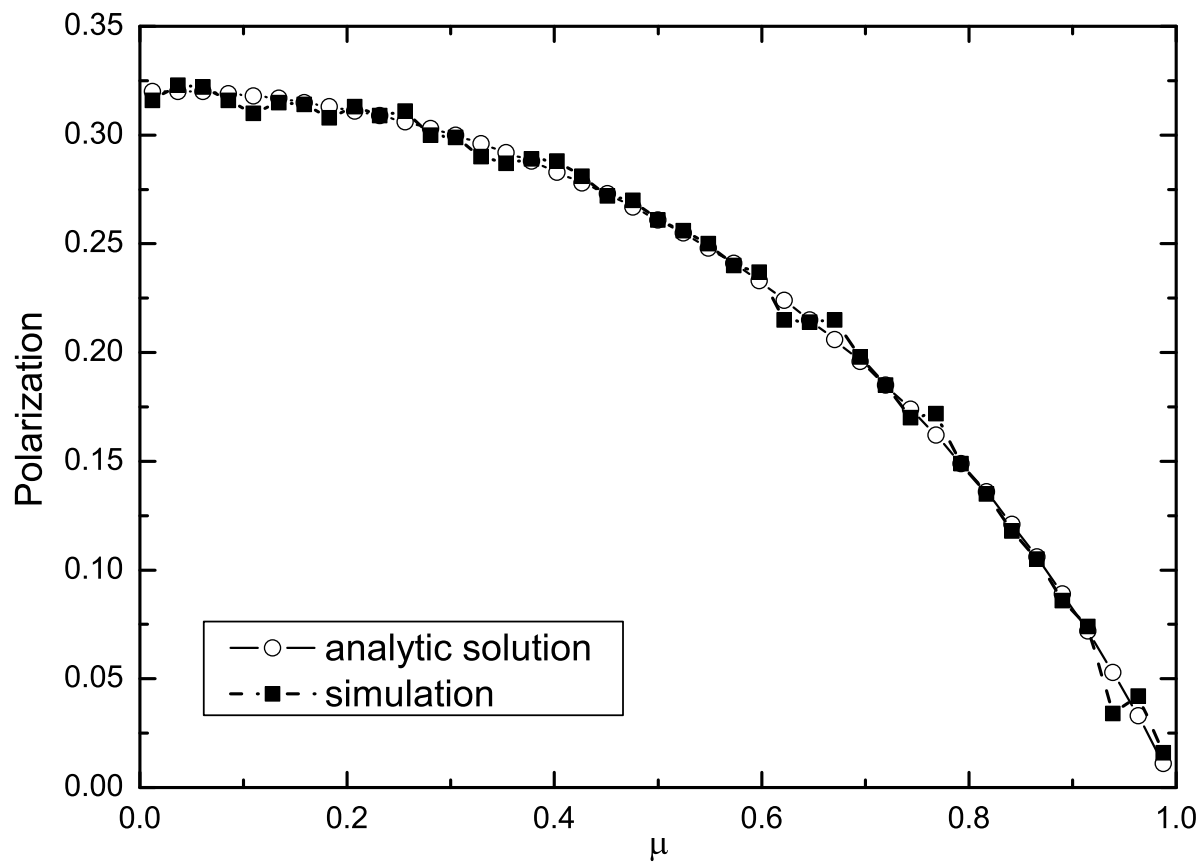}
  \caption{The output polarization of the single-scattered light viewed at different inclination angle $i$ in the case of optically thin limit ($\tau << 1$). The geometry of the scatterer is
presented in Fig. \ref{model}. Here $\mu=\cos i$,
$\theta_0=12^{\rm o}$, and $N_e$ = $4\times10^{23}$ cm$^{-2}$. The
analytic result is presented as open circles (eq. 7), and the
simulation resuls as black squares.} \label{single-cs}
\end{figure}

\begin{figure}
\plotone{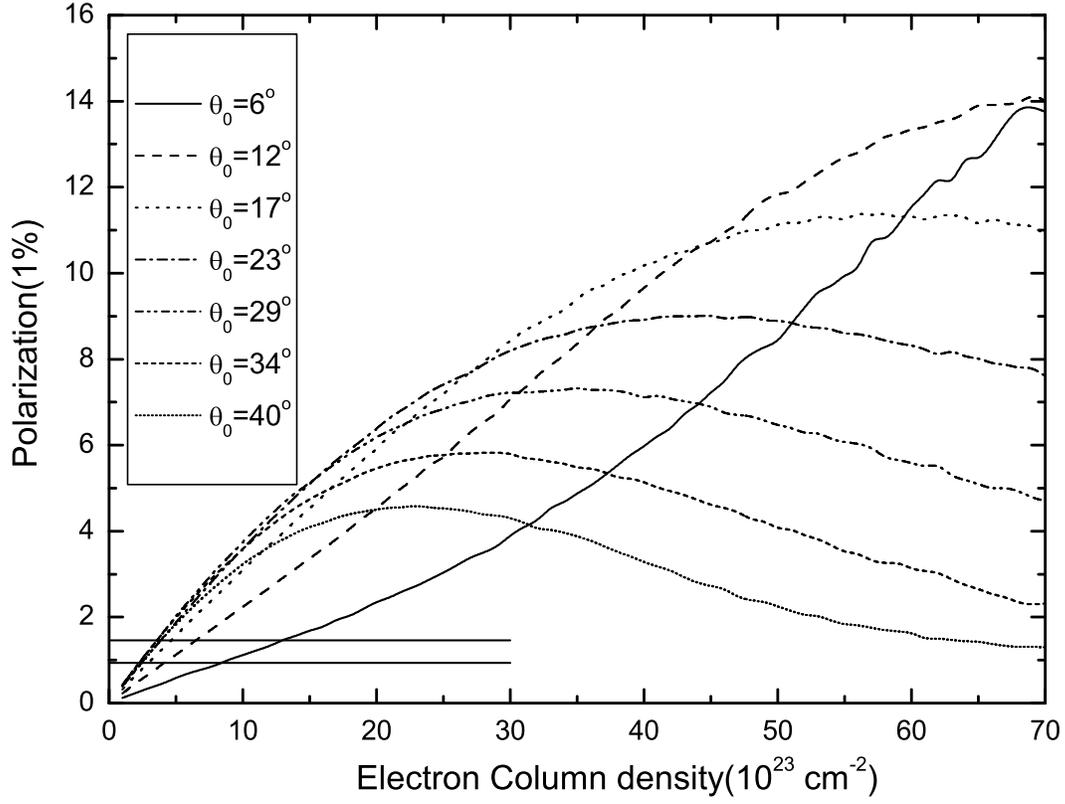} \caption{The average continuum polarization
degree for BAL QSO as a function of the column density for
different $\theta_0$, based on Monte-Carlo simulations. The two
horizontal lines mark the polarization degree 0.93\% (the average
value of HiBAL QSOs) and 1.46\% (the average of LiBAL QSOs). The
data }\label{pbar-BAL}
\end{figure}

\begin{figure}
\plotone{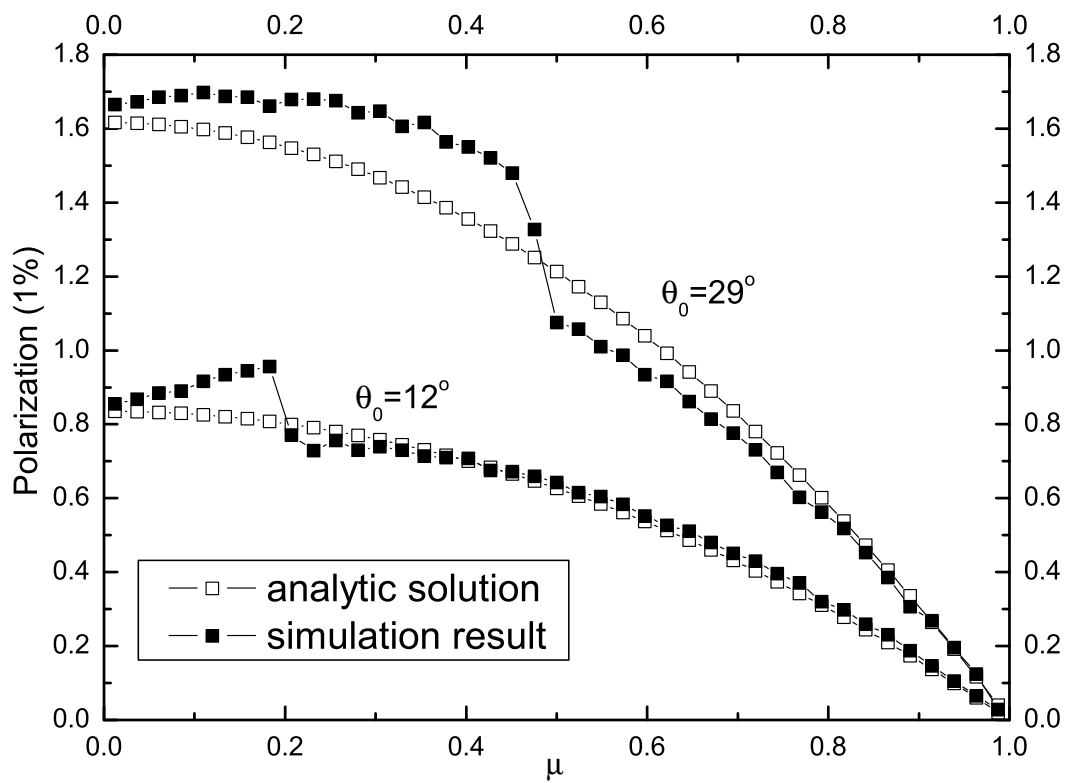} \caption{The effect of the radiation transfer on
the output polarization for $N_e=4\times10^{23}$ cm$^{-2}$
($\tau_e=0.266$). The analytic result takes into account of only
single scattering while both multiple scattering and the
attenuation of the direct light is considered in the Monte-Carlo
simulation. }\label{transfer}
\end{figure}

Following the procedures described in \S2, we perform Monte-Carlo
simulations to correct the effect of radiative transfer. We test
the Monte-Carlo code by simulating single scattering process, and
found the results from the simulations are in good agreement with
the analytic ones (Fig. \ref{single-cs}). In the simulations
below, we consider the half covering angle $\theta_0$ from 6$^{\rm
o}$ to 40$^{\rm o}$ and the column density of electron $N_e$ from
$10^{23}$~cm$^{-2}$ to $7\times10^{24}$~cm$^{-2}$. In Fig.
\ref{pbar-BAL} we plot the resulted average continuum polarization
degree ($\bar{P}_B$) for BAL QSO as a function of the column
density. We can see that, to produce the observed average
polarization degrees of HiBAL (0.93\%), the minimum column density
of 2~$\times 10^{23}$~cm$^{-2}$ is required for the range of
$\theta_0$ considered. This number increases to 3.5~$\times
10^{23}$~cm$^{-2}$ for LiBAL. Using the average $\theta_0=12^{\rm
o}$ (corresponding to $\mu_0$=0.2), we derive a column density of
4~$\times 10^{23}$~cm$^{-2}$ for HiBAL, slightly higher than the
analytical result. In Fig. \ref{transfer}, we plot the output
polarization degree as a function of the viewing angle from Monte
Carlo simulations, assuming $N_e$ = 4 $\times 10^{23}$ cm$^{-2}$
($\tau_e$ = 0.266). For comparison, the analytic results, which is
valid when $\tau_e$ $<<$ 1, are also presented. The impact of the
radiative transfer to the polarization degree is apparent in the
figure. Basically, because of the attenuation of the direct
continuum, the radiative thransfer causes larger polariztion for
BAL QSOs than that from the analytic calculation. For non-BAL
QSOs, the situation is more complex. For a model with small
$\theta_0$ = $12^{\rm o}$, the optical depth at small inclination
angle is smaller, consequently the scattered photons escape more
likely along the polar direction, which results in slightly larger
(smaller) polariztion at smaller (larger) inclinations. However,
for a model with larger $\theta_0$ = $29^{\rm o}$, the attenuation
of the scattered light becomes important, which reduces the
polarization for non-BAL QSOs.

\begin{figure}
\epsscale{0.45}\plotone{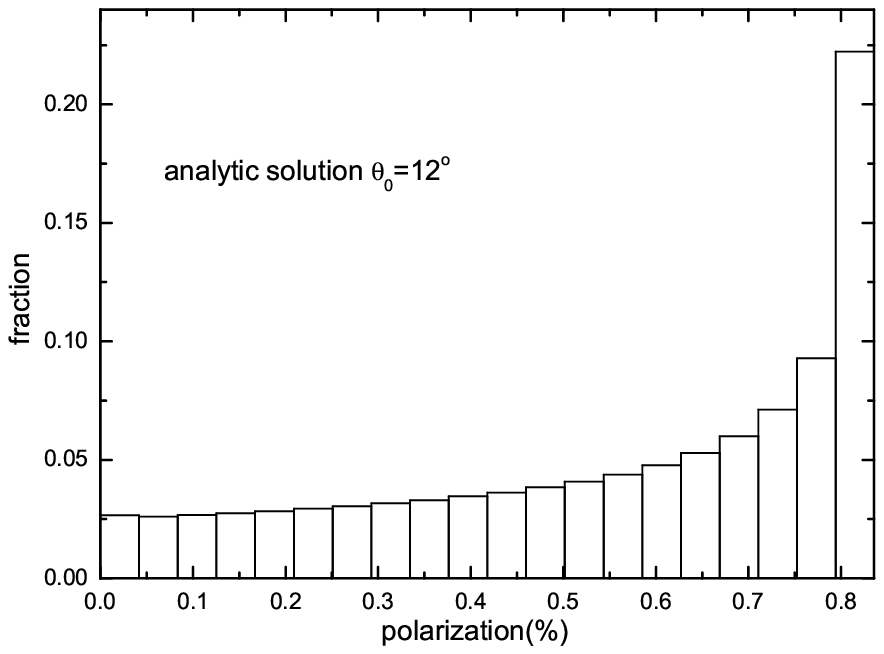}\plotone{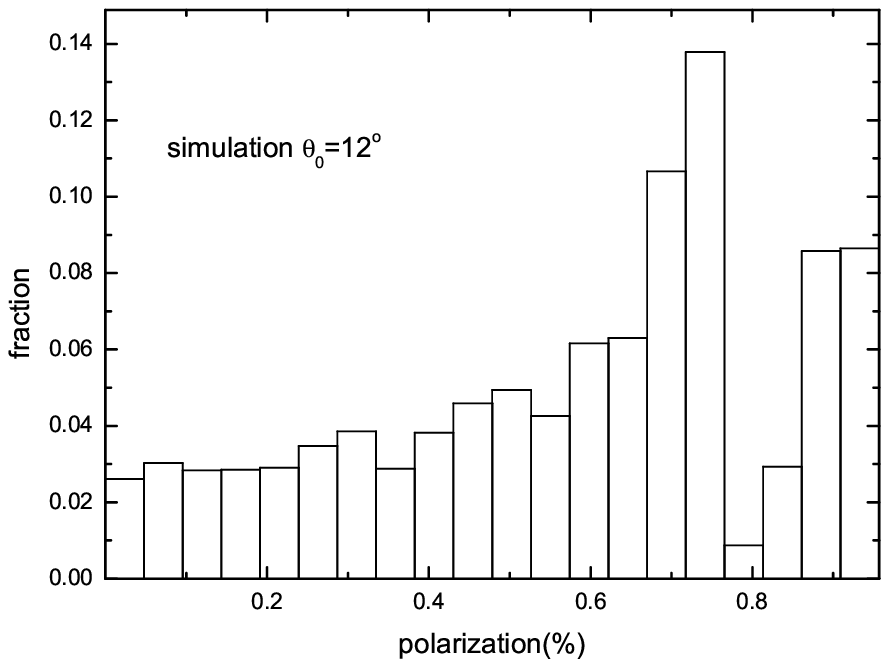}
\plotone{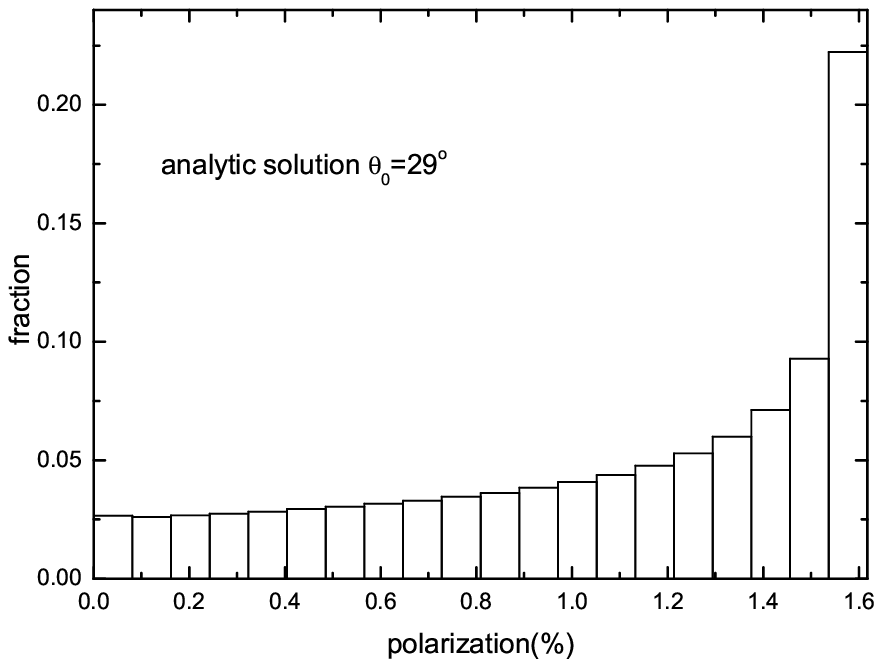}\plotone{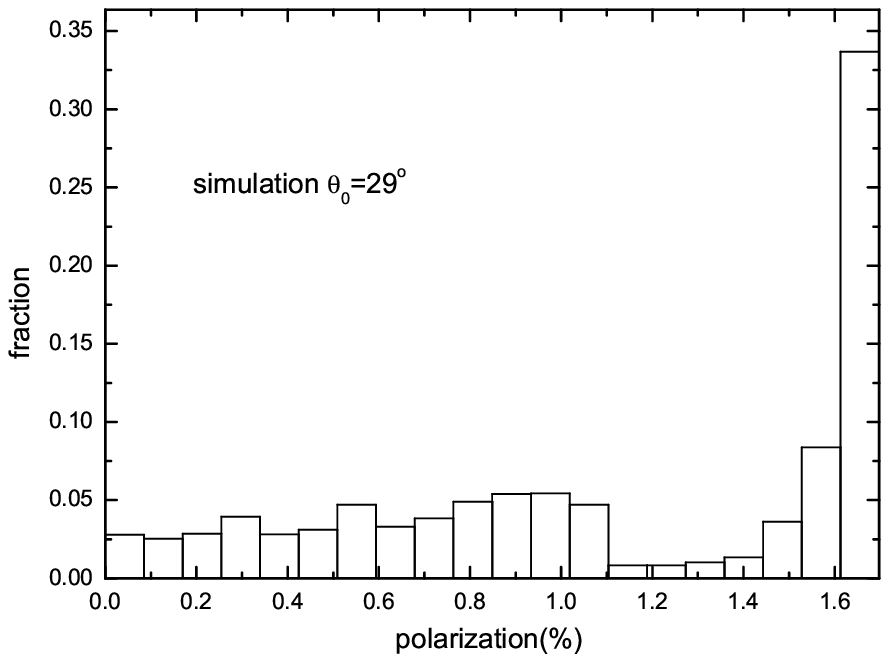} \caption{The distributions of
the polarization degree for models with $\theta_0=12^{\rm o}$ and
$\theta_0=29^{\rm o}$. The left two panels are the analytic
results and the two right panels are the simulation results. A
column desity $N_e=4\times10^{23}$ cm$^{-2}$ is assumed.
}\label{fdp0205}
\end{figure}

We point out that, in additional to the observed average
polarization degrees for BAL and non-BAL QSOs, the distribution of
the observed polarization of a radio quiet QSO sample can provide
further constraints to the model of the scatterer.
We consider the optically thin case first. For a given
axisymmetric electron-scatterer, the polarization degrees at
different view angles $i$ follow eq. \ref{eq:p_e}.
The normalized distribution of the polarization degree can thus be
written as

\begin{equation}\label{eq:dndp1}
\frac{dN}{dp}=\frac{1}{dp/d\mu}=\frac{1}{2p_0\sqrt{1-p/p_0}}\;\;
{\mathrm{for}}\; p\le p_0
\end{equation}

where $p_0$ is the polarization degree viewed at the direction
perpendicular to the symmetric axis. Note $p_0$ depends on the
optical depth and the geometry of the scatterer. We notice that
the distribution is a monotonic increasing function of $p$ for a
given scatterer. Monte-Carlo simulations, which taking account of
radiative transfer for larger optical depth, give similar result
(see. Fig. \ref{fdp0205}). However, an opposite trend was found in
the observed distributions given by Stockman et al (1984) and
Berriman et al. (1990), who found the observed distribution of
polarization degree of all QSOs (including BAL QSOs) decreases
with increasing $p$ above 0.2\%. We argue that the discrepancy
might also be explained by a dispersal distribution of the
properties of the shielding gas. Here considering a distribution
of $p_0$ as $f(p_0)$, we can write formally the distribution of
$p$ as

\begin{equation}
\frac{dN}{dp}=\int_p^1 \frac{f(p_0)dp_0}{2p_0\sqrt{1-p/p_0}}
\end{equation}

\begin{figure}
\epsscale{1.0}\plotone{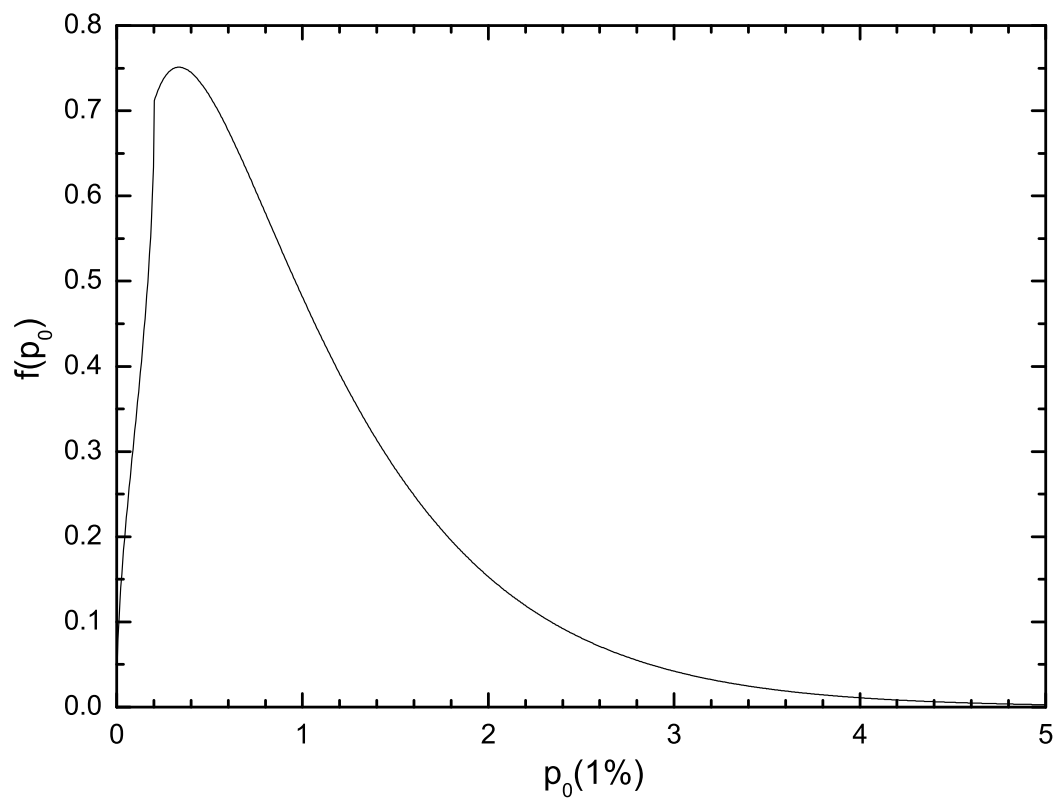} \caption{The reconstructed $p_0$
distribution assuming an axisymmetric electron-scattering model,
where $p_0$ is the polarization degree detected on equatorial
plane. See text for details. }\label{fp0}
\end{figure}

If the observed distribution of the polarization is known, we can
solve the above equation inversely, e.g.,through Richardson-Lucy
approach, to obtain the $f(p_0)$, which in turn can be used to
constrain the geometry of the scatterer. Using the polarization
distribution of QSOs presented by Stockman et al (1984, see Fig. 1
in their paper), which can be described approximately with

\[ \frac{dN}{dp}=\left\{
           \begin{array}{ll}
               16.5 &  (p<0.2\%) \\
               22.23\exp(-p/0.0067) & (p>0.2\%)
            \end{array}
            \right.
\]

we obtain $f(p_0)$ in Fig. \ref{fp0}, which rises steeply toward
lower $p_0$ with a peak at $p_0\simeq0.34\%$. This suggests that
the shielding gas in a large fraction of QSOs either have smaller
column density, or smaller covering factor. Note that $P$ =
$0.34\%$ is much smaller than the average continuum polarization
of BAL QSOs which is 0.93\%. This suggests that most of the
quasars have BALR with covering factor much smaller than 0.2, thus
these BALR make less contribution to the observed sample of BAL
QSOs because of the smaller chance to be detected. A larger sample
of radio quiet quasars with accurate measurement of the continuum
polarization will help give better constraints to the distribution
of the BALR properties.  Note the observed distribution at the
small polarization end ($p<0.2\%$) was not well constrained
because of the limited accuracy of the measurements, thus the
resulted shape of $f(p_0)$ has large uncertainty.

Note that the derived $f(p_0)$ is also valid for axisymmetric
scattering models, including the polar scattering models (e.g.,
Ogle 1997; Lamy \& Hutsem{\'e}kers 2004) and possible bi-conical
scattering model (Elvis 2000). The reason is that the polarization
produced by axisymmetric outflow is $\propto\sin^2i$, but
independent to the outflow model (eq.\ref{eq:p_e}).

In passing, we point out that the average level, as well as the
angle-dependence, of the polarization degree may be affected by
the anisotropic emission of the continuum source, which is
neglected in the above calculations. Certain levels of anisotropy
are expected in optically thick accretion disk models, which
usually predict stronger emission in polar directions than on the
equatorial plane. In such models, the average polarization degree
would be lowered in our equatorial scatterer model, and the
polarization degree increases with inclination faster than in the
case of isotropic continuum emission. However, we believe that the
anisotropic emission would not severely affect our results for two
reasons. First, BAL QSOs are at least as luminous as non-BAL QSOs
\footnote{In fact, Boroson et al. (2002) found that BAL QSOs are
observed more luminous than the non-BAL QSOs for a given mass of
black hole.}. Second, we do not find anti-correlation between the
polarization degree and the continuum luminosity (Moore, R. L. \&
Stockman, H. S. 1984), which is expected in a model with
anisotropic continuum.

\section{Conclusion}

We show that the X-ray absorber in the BAL QSOs is capable to
reproduce the observed continuum polarization for both BAL QSOs
and non-BAL QSOs. However, the covering factor of the BALR in
quasars is required to have a dispersal distribution, instead of a
$\delta$ function. We also find that to macth the observed
distribution of the continuum polarization of radio quiet quasars,
the BALR in most QSOs produces much smaller maximum continuum
polarization $p0$ (while viewed edge-on) with a peak at $p0$ =
0.34\%, which is much smaller than the average continuum
polarization of BAL QSOs, which is 0.93\%. Consequently, the BAL
QSOs with small $p0$ are likely to have covering factor of BALR
much smaller than 0.2, thus make less contribution to the observed
sample of BAL QSOs because of the smaller chance to be detected.

\end{document}